\documentclass[preprint2]{aastex}
\def\kms{\rm km\ s^{-1}}

\def\kpch{\mbox{$h^{-1}$kpc}}

\def\hmpc{\mbox{$h^{-1}$Mpc}}
\def\Mpch{\mbox{$h^{-1}$Mpc}}

\def\Msunh{\mbox{$h^{-1}$M$_\odot$}}

\def\LCDM{{$\Lambda$CDM}}
\def\m3{MARK III}
\def\br{{\bf r}}
\def\bv{{\bf v}}
\def\be{\begin{equation}}
\def\ee{\end{equation}}
\def\CR{^{\rm CR}}

\def\mathnew{\mathsurround=0pt}
\def\simov#1#2{\lower .5pt\vbox{\baselineskip0pt
    \lineskip-.5pt\ialign{$\mathnew#1\hfil##\hfil$\crcr#2\crcr\sim\crcr}}}

\def\'#1{\ifx#1i{\accent"13\i}\else{\accent"13#1}\fi}

\shorttitle{Simulations of the Real Universe}
\shortauthors{Klypin et al.}
\begin{document}
\title{Constrained Simulations of the Real Universe: the Local
Supercluster}

\author{Anatoly Klypin}
\affil{Astronomy Department, New Mexico State University, Las Cruces, NM
88003-0001}
\author{Yehuda Hoffman}
\affil{Racah Institute of Physics, Hebrew University, Jerusalem
91904, Israel}
\author{Andrey V. Kravtsov\footnote{Hubble Fellow}}
\affil{Department of Astronomy, Ohio State University, 140 West 18th Ave.,
Columbus, OH 43210-1173}
\author{Stefan Gottl\"ober}
\affil{Astrophysikalisches Institut Potsdam, An der Sternwarte 16, D-14482
Potsdam, Germany}

\begin{abstract}  

  We present cosmological simulations which closely mimic the real
  Universe within $\sim 100$~Mpc of the Local Group.  The simulations,
  called Constrained Simulations, reproduce the large-scale density
  field with major nearby structures, including the Local Group, the
  Coma and Virgo clusters, the Great Attractor, the Perseus-Pices, and
  the Local Supercluster, in approximately correct locations.  The
  \m3\ survey of peculiar velocities of the observed structures inside
  $80\hmpc$ sphere is used to constrain the initial conditions.
  Fourier modes on scales larger then a (Gaussian) resolution of
  $\approx R_g=5 \hmpc$ are dominated by the constraints, while small
  scale waves are essentially random.  The main aim of this paper is
  the structure of the Local Supercluster region (LSC; $\sim 30\hmpc$
  around the Virgo cluster) and the Local Group environment.  We find
  that at the current epoch most of the mass ($\approx 7.5\times
  10^{14}\Msunh$) of the LSC is located in a filament roughly centered
  on the Virgo cluster and extending over $\sim 40 \hmpc$.  The
  simulated Local Group (LG) is located in an adjacent smaller
  filament, which is not a part of the main body of the LSC, and has a
  peculiar velocity of $\approx 250~\kms$ toward the Virgo cluster.
  The peculiar velocity field in the LSC region is complicated and is
  drastically different from the field assumed in the Virgocentric
  infall models.  We find that the peculiar velocity flow in the
  vicinity of the LG in the simulation is relatively ``cold'': the
  peculiar line-of-sight velocity dispersion within $7 \hmpc$ of the
  LG is $\lesssim 60~\kms$, comparable to the observed velocity
  dispersion of nearby galaxies.

\end{abstract}
\keywords{cosmology:theory -- large-scale structure
-- methods: numerical}


\section{Introduction}
\label{sec:intro}

Over the last two decades cosmological simulations have proved to be
an invaluable tool in testing theoretical models in the highly
nonlinear regime. However, comparisons of simulations with
observational data are typically done only in a statistical sense.
The standard approach is to assume a cosmological model and to use the
appropriate power spectrum of the primordial perturbations to
construct a random realization of the density field within a given
simulation volume.  The evolution of the initial density field is then
followed using a numerical code and results are compared with
observations. A wide variety of statistical measures has been used for
comparisons, including the two-point correlation function, the power
spectrum, the mass function, different shape statistics, etc.

The statistical approach works well if there is a statistically
representative sample of objects with well understood selection
effects for both the observed Universe and the simulations. The
number of sample objects should be sufficient to overcome the cosmic
variance. The approach fails, however, in cases dealing with rare
objects such as the Great Attractor or specific configurations as, for
example, in the problem of the local peculiar velocity field 
\citep{peebles92}. The traditional solution has been to choose
``observers'' or regions in the simulations that resemble the desired
configuration as closely as possible. The results of numerical
simulations aimed at identifying structures similar to those observed
locally in random realizations of the initial density field were
generally inconclusive
because one is never sure that the selection of objects in simulations is 
right.

The catalog of galaxy groups constructed using the CfA redshift survey
\citep{davis_etal82} is a good illustration of the difficulties one
encounters in comparing model predictions to observations
\citep{huchra_geller82, nolthenius_etal97}. While the data-set was
sufficiently large for statistical comparisons with models, a large
fraction of groups in the sample is in 
one large object: in the Local Supercluster (LSC). 
For detailed comparison with observational data it was important to
identify groups in a similar environment in cosmological simulations.
For example,  it was crucial to mimic the
environment of the Local Group (LG) in the LSC
\citep[e.g.,][]{nolthenius_etal97}. This was done by simply choosing
an ``observer'' $20\hmpc$ away from a cluster with mass comparable to
that of the Virgo Cluster. Unfortunately, it is not clear what
``similar environment'' actually means and that simply placing the
``observer'' at some distance from ``Virgo'' cluster resolves the
issue.

Another problem that is not easy to address using statistical
comparisons is the peculiar velocity field around the LG. The
observed low velocities of nearby galaxies have long being perceived
as one of the important challenges for structure formation models
\citep{peebles92}. Because we deal with rather specific environment,
 in simulations we must find objects, which are analogous to the real Local
Group. The interpretation of the galaxy velocity field is further
complicated by the possibility of velocity bias
\citep{gelb_bert94,colin_etal00}. Nevertheless, it was suggested that,
even after taking the uncertainties into account, observed peculiar
velocities of galaxies around the LG are unusually low if
compared to the typical velocities of galaxies in cosmological
simulations \citep{schlegel_etal94,governato_etal97}.

There are two possible ways of addressing problems, which require
comparison with specific astronomical objects or with specific
environments. ({\bf i}) We may simply ignore them and wait for larger
observational data-sets, such as the Sloan Digital Sky Survey, and
pursue the statistical comparison of the observed data with the
theoretical predictions.  Clearly, the increasing amount and quality
of data will continue to fuel the statistical way of testing the
cosmological models. At the same time, observational data always will
be more extensive and more complete for nearby, somewhat unique
objects. Thus, by doing only statistical tests we are bound to lose
information on the best observed astronomical objects in our sample.

({\bf ii}) An alternative approach, adopted in the present study, is
to find a way of making  simulations, which  reproduce the large
scale structures of the real Universe, while keeping the initial
conditions consistent with the power spectrum of a given
cosmological model.  In particular, we want to simulate the small
scale nearby structures  within the correct large environment
of the real Universe as revealed by various large scale
surveys.  This is done by setting initial conditions of the
simulations by constrained realizations of Gaussian fields
\citep{hoffman_ribak91}, thereby constructing the density and velocity
fields which agree both with the observed large scale structures and with
the assumed theoretical model.  This approach is called
here as constrained simulation (CS).

Low resolution CSs were made before  by \citet{kolatt_etal96} 
and  \citet{bistolas_hoffman98}, using the IRAS 1.2Jy redshift survey 
to set the constraints, with the main aim of producing a 
semi-realistic non-linear realization of the large scale structure.
Recently, such CSs were made using the \m3\ survey of radial 
velocities \citep[][]{willick_etal97}, and were used \citep{wh99,wh00} 
to study the origin of the local cold velocity field.  The study 
presented here extends this work by using numerical simulations of 
much higher spatial and mass dynamic range.  The dynamic range 
achieved by the Adaptive Mesh Refinement simulations used in this 
study far exceeds the range of all previous CSs.  The combination of 
CSs and adaptive mesh refinement is the optimal strategy for 
simulating formation of the local structures at high spatial and mass 
resolution in a computational box sufficiently large to properly model 
the large-scale tidal field.

The present work is the first in a series of papers studying the
dynamics of the nearby universe by means of CSs.  The main purpose of
this paper is to present the general approach and details of the
method.  The specific CS analyzed here is a relatively low resolution
one, while forthcoming CSs will use higher resolution initial
conditions and will reach higher mass and spatial dynamic range. As an
illustration, this paper addresses the question of morphology of the
density field and dynamics in the LSC region.  In the following
sections we present analysis of the structure of the Local
Supercluster, as well as the smoothness of the Hubble expansion and
peculiar velocity field in the simulated LSC region. We will also
discuss the accuracy of the Virgocentric model in describing the
velocity field in our simulation.  The upcoming studies will focus on
a variety of problems, including gas-dynamic simulations of the Virgo
cluster and the LSC region \citep{kkh01},
high-resolution simulation of the LG, halo/galaxy formation
in the Local Void region, and biased galaxy formation in the local
universe.

The paper is organized as follows. A brief review of studies of the
LSC region is given in \S~\ref{sec:lsc}.  The formalism of constrained
realizations, the \m3\ survey, and the construction of the initial
conditions are presented in \S~\ref{subsec:WF/CR},
\S~\ref{subsec:data}, and \S~\ref{subsec:applic}, respectively.  The
details of the numerical simulation are described in
\S~\ref{subsec:CS} and \S~\ref{sec:numer}.  The analysis of the CS is
presented in \S \ref{sec:results} and we conclude the paper with a
discussion of the results in \S~\ref{sec:disc}.

\section{The Local Supercluster}
\label{sec:lsc}

Ever since the first large samples of nebulae compiled by William and
John Herschel, it was known that there is a marked excess of bright
objects in the northern hemisphere around the Virgo cluster, with most
nebulae concentrated in a wide band spanning some $100^{\circ}$
\citep[e.g.,][]{reynolds23,lundmark27,shapley34}\footnote{See
interesting historical note on ``discovery'' of the LSC by
\citet{dev89} and \S~2 in historical review by \citet{biviano01}}.  In
1950s \citet{dev53,dev58} was the first to argue that this excess
corresponds to a real 3-dimensional structure in galaxy distribution.
At first \citet{dev53} called the structure the ``Supergalaxy'' (and
introduced the supergalactic coordinates in analogy with galactic
coordinate system) conjecturing that the apparent flattened galaxy
distribution represented a large-scale galaxy-like system. later he
subsequently changed the name to the ``Local Supercluster''
\citep[LSC;][]{dev58}, which has been used in the literature ever
since.

The debate on whether the LSC is a physical system or a chance
alignment of galaxies has continued over several decades
\citep[e.g.,][]{bahcall_joss76,dev76,dev78} and was settled with
the advent of large redshift surveys of nearby galaxies
\citep{sandage_tamman81,fisher_tully81,davis_etal82}.
\citet{yahil_etal80} and \citet{tully82}, for example, analyzed the
morphology of the 3-dimensional distribution of galaxies in the
\citet{sandage_tamman81} and \citet{fisher_tully81} surveys,
respectively, and showed convincingly that the concentration of
galaxies along the supergalactic plane on the sky does indeed
correspond to a flattened large-scale structure in the distribution of
nearby galaxies. \citet{einasto_etal84} applied a battery of
tools to quantify the statistical properties of the galaxy distribution in
the LSC region and compared the results with quantitative and qualitative
predictions of several cosmological models. In particular, they showed
that the LSC is similar in structure and morphology to other known
superclusters such as the Perseus-Pisces supercluster and is a part of
the large-scale network of clusters, sheets, and filaments.

More recent studies \citep{tully_fisher87,km96,lahav_etal00} showed
that the main body of the LSC is a filamentary
structure that extends over some $40\hmpc$ and is roughly centered on
the Virgo cluster. The whole region is dominated by several clusters
(Virgo, Ursa Major, and Fornax are the most prominent), groups,
and filaments, the latter bordering nearby voids (such as the Local
Void). The LG is located in the outskirts of this region in a
small filament extending from the Fornax cluster to the Virgo
cluster \citep{km96}.  The LSC contains a fair number of galaxy groups
\citep{huchra_geller82} and its skeletal structure is well traced by
radio galaxies and AGNs \citep{shaver_pierre89,shaver91}.

The quest for the Hubble constant and the advent of peculiar velocity
surveys of galaxies fueled studies of galaxy dynamics in the LSC
region \citep[see, e.g.,][ and references
therein]{davis_peebles83,huchra88}.  In particular, it was recognized
that the infall of the LG toward the LSC could be used to measure the
mass-to-light ratio (and hence the matter density in the Universe,
$\Omega_0$) on large scales \citep{silk74,peebles76}. The Virgocentric
infall model \citep{gunn78} was used to estimate $\Omega_0\sim
0.2-0.3$ from the observed local peculiar flow of galaxies
\citep{huchra88}. This model requires rather large non-linear
corrections \citep{yahil85,villumsen_davis86} and was shown to be
quite inaccurate when tested using outputs of numerical cosmological
simulations \citep{villumsen_davis86, Lee_Hoffman_Ftaclas_86,cen94,
governato_etal97}, rendering the conclusions about $\Omega_0$
inconclusive.

The quiescence of the local peculiar velocity field
\citep{dev58,sandage_tammann75,rivolo_yahil81,
sandage86,giraud86,schlegel_etal94,km96,sandage99,ekholm_etal01} is a
long standing puzzle which presents a challenge for models of
structure formation \citep[e.g.,][]{peebles92}. That this puzzle is by
no means solved is clear from the recent study by \citet{sandage99}
who comments that ``the explanation of why the local expansion field
is so noiseless remains a mystery.'' The radial peculiar velocity
dispersion of galaxies within $\approx 5\hmpc$ around the Milky Way is
only $\sim 50-60~\kms$ and the local Hubble flow agrees with the
global expansion law on large scales to better than $10\%$
\citep{sandage99}. A striking manifestation of the coldness of the
flow is the fact that outside the Local Group there are no nearby
galaxies with blueshifts with respect to the Milky Way.

Clearly, successful cosmological models should provide a plausible
explanation for such a ``cold'' velocity field. However, we are dealing
with a unique and limited set of data which makes it difficult to employ
the usual statistical comparisons between models and observations.
Theoretical studies employing cosmological simulations concluded that
the CDM models fail to naturally reproduce the ``coldness'' of the
local velocity field \citep{schlegel_etal94,governato_etal97}.
\citet{schlegel_etal94} used a variety of criteria to identify LG
candidates in their simulations of the standard CDM and mixed DM (MDM,
or cold$+$hot dark matter) models.  They found that the observed
``coldness'' of the local flow cannot be reproduced in the then
standard CDM model, but could be plausibly reproduced in the MDM
model.  \citet{governato_etal97} used a set of somewhat different
criteria to identify LG counterparts in the simulations of the
standard and open CDM models.  They selected binary systems consisting
of halos with circular velocities of $125- 270~\kms$ approaching each
other, which do not have other massive neighbors within 3~Mpc. An
additional subsample was constructed by requiring that the selected
binary systems are located at $5-12\hmpc$ from a Virgo sized cluster.
This study confirmed that the standard CDM model cannot reproduce the
observed low value of velocity dispersion.  The velocity dispersions
around groups in the open CDM model with mass density $\Omega_0=0.3$
($\sim 150-300~\kms$) were found to be lower than those in the SCDM
($\sim 300-700~\kms$), but still higher than the observed value.  The
authors concluded that ``neither the $\Omega=1$ (CDM) nor $\Omega=0.3$
(OCDM) cold dark models can produce a single candidate LG
that is embedded in a region with such small peculiar velocities.''

The conclusions of these studies, however, have to be viewed with caution
as they explored only a limited range of cosmological models (in
particular, the currently favored {\LCDM} model was not considered)
and have used a statistical approach to select possible LG
candidates from a fairly small-size simulation volume.  For example,
it is not clear whether simple criteria lead to selection of objects
representative of the LG or, if the LG environment is rare,
that the box size was sufficient to contain a suitable counterpart.
It is also not clear what is the largest scale that influences the
specific properties of the LG environment.  The local ``cold'' flow
may be influenced by the mass distribution on much larger scales and
be induced by coupling of the small and large scales {\it via} the
velocity shear \citep[see ][]{wh99,wh00}. The local criteria for the
definition of LG counterparts are then clearly inadequate.


\section{Constrained Simulations}
\label{sec:CS}

\subsection{Wiener Filter and Constrained Realizations of Gaussian Fields}
\label{subsec:WF/CR}

The standard cosmogonical framework of the CDM family of models
assumes that structures evolve out of small perturbations in an
expanding Friedmann universe. It is assumed that these perturbations
constitute a Gaussian random field. Redshift and radial velocity
surveys provide information that enables the reconstruction of the
large-scale structures (LSS) in the density field of the nearby
Universe. An efficient algorithm for reconstructing the density and
velocity fields from sparse and noisy observations of the LSS, such as
the one provided by redshift and velocity surveys, is provided by the
formalism of the Wiener filter \citep[WF; for a general overview see
][]{zaroubi_etal95}.  The application of the WF requires some model
for the power spectrum that defines the statistical properties of the
perturbation field.  A Bayesian approach to the problem of choosing
the ``best'' model consists of finding the most probable model
over an assumed parameter and/or model space by means of maximum
likelihood analysis, and using this model for calculating the WF (see,
however, \citealt{hoffman_zaroubi00} for a discussion of the
limitations of the Bayesian approach.) The application of the general
formalism to the case of radial velocity surveys follows
\citet{zaroubi_etal97} and \citet{zaroubi_etal99}.

Below we briefly describe the WF reconstruction method. Consider a survey of
$N$ objects with radial velocities $\{ u_{i}\}_{i=1,\ldots,N}$, where
\begin{equation}
        u_{i} = {\bf  v} ({\bf r}_{i})  \cdot \hat {\bf r}_{i}
                +\epsilon_{i}, \label{eq:ui}
\end{equation}
\noindent Here ${\bf  v}$ is the three dimensional velocity, ${\bf r}_i$ is the
position of the i-th data point,  and $\epsilon_{i}$ is the statistical
error associated with the $i$-th radial velocity. Given that the assumed
model, and that the statistical errors are well understood, the data
auto-covariance matrix can be readily evaluated:
\begin{equation}
         \Bigl < u_i u_j  \Bigr > =  \hat {\bf r}_j \Bigl < {\bf  v}
                ({\bf r}_i) {\bf  v} ({\bf r}_j)  \Bigr >
                   \hat {\bf r}_j   + \sigma{^2_{ij}}.
\label{eq:Rij}
\end{equation}
Here $\Bigl <  \ldots  \Bigr >$ denotes an ensemble average.
The last term $\sigma^2_{ij}$ is the error covariance matrix.
The velocity covariance tensor is calculated using linear theory.

Given a peculiar velocity dataset, the WF provides the minimum
variance estimation of the underlying field(s). In the case where the
underlying field is Gaussian, the WF-reconstructed density field
coincides with the most probable and the mean fields for the data, and
also with the maximum entropy estimation \citep{zaroubi_etal95}.
Assuming the linear theory relation of the radial velocity with the
full three-dimensional velocity and density fields, the WF estimation
of these fields is:
\begin{equation}
{\bf v}^{WF}({\bf r}) =
        \Bigl < {\bf v}({\bf r}) u_i \Bigr >
                 \Bigl < u_i u_j \Bigr >^{-1} u_j \label{eq:WFv}
\end{equation}
and
\begin{equation}
        \delta^{WF}({\bf r}) = \Bigl < \delta({\bf r}) u_i \Bigr >
                  \Bigl < u_i u_j\Bigr > ^{-1}    u_j\label{eq:WFd}
\end{equation}
The $\Bigl < {\bf v}({\bf r}) u_i \Bigr >$ and
$\Bigl < \delta({\bf r}) u_i \Bigr >$ are the cross radial
velocity -- three-dimensional velocity and density correlation matrix.

The WF is a very conservative estimator.  In the absence of ``good''
data, namely where the data is sparse and/or noisy, it attenuates the
estimate towards its unbiased mean field, which in the cosmological
case is the null field.  Thus, by construction the WF suppresses some
of the power that is otherwise predicted by the assumed model.  The WF
often produces an estimated field that is much smoother than the typical
random realization of the assumed power spectrum would be.  In
particular, the WF estimator is not statistically homogeneous.  A way
of providing the missing power and regaining the statistical
homogeneity consistent with the data and with the theoretical model is
provided by the method of constrained realizations of Gaussian fields
\citep{bert87,hoffman_ribak91,hoffman_ribak92}.  The constrained realizations provide a
realization of the underlying field made of two components.  One is
dictated by the data and by the model and the other is random in such
a way that in places, where the WF suppresses the signal, the random
component compensates for it.

The WF attenuation can be overcome by the unbiased minimum variance
estimator of \citet{zaroubi01}, where some of the missing power is
provided by the observational errors. It provides an attractive tool
for recovering the LSS, but it is not suitable for providing
realizations that are consistent with a theoretical model.

The \citet{hoffman_ribak91} algorithm of constrained realizations provides a very
efficient way of creating typical realizations of the residual from
the WF mean field.  The method is based on creating random
realizations of the density and velocity fields, $\tilde\delta(\br)$
and $\tilde\bv(\br)$, given an assumed power spectrum and a proper set of
random errors $\tilde\epsilon_i$.  The  random realization is
then ``observed'' just like the actual data to yield a mock velocity data set
$\tilde u_i$.  Constrained realizations of the dynamical fields are
then obtained by

\be
        \bv \CR(\br) = \tilde \bv(\br) + \Bigl < \bv(\br)
        u_i \Bigr > \Bigl < u_i u_j \Bigr > ^{-1} \bigl( u_j -\tilde u_j\bigr)
\label{eq:CRv}
\ee
and
\be
        \delta \CR(\br) =\tilde \delta(\br) +
                \Bigl < \delta(\br) u_i \Bigr >  \Bigl < u_i u_j\Bigr > ^{-1}
                 \bigl( u_j -\tilde u_j \bigr) .\label{eq:CRd}
\ee

The variance of the constrained realizations around the WF mean field provides a measure of
the amount by which they are constrained by the data. This variance
can be calculated rigorously using the auto- and cross-correlation
matrices defined here \citep{zaroubi_etal95}. However this calculation
becomes impractical for large grids as it involves the inversion of
large matrices. A much simpler and more efficient way is to construct
an ensemble of constrained realizations and to calculate its scatter 
\citep{zaroubi_etal99}.

The auto- and cross--covariance matrices in the above equations are
computed within the framework of the linear theory. The WF and the
constrained realizations are performed assuming that both the sampled
data and the evaluated fields are in the linear regime. Indeed, for
velocity data processed by the grouping algorithm, the linear theory
provides a good approximation \citep[see][]{kudlicki_etal01}.  The
choice of the resolution at which the WF and fields of constrained
realizations are evaluated is arbitrary and it can be controlled by a
Gaussian smoothing of radius $R_g$ and of course by the grid scale
over which the fields are constructed.  Besides the practical
limitations imposed on choosing the grid size, there is one basic
consideration that dictates the choice of the grid and the
smoothing. This can be best understood in terms of the Fourier space
presentation, where in the case of unconstrained realizations the
Fourier modes are statistically independent. The imposed constraints
introduce mode coupling, where the range of the coupling depends on
the kind of data (say, density or velocity), its quality (errors,
sparseness and space coverage) and the nature of the power spectrum
\citep{bert87,hoffman_ribak91,hoffman_ribak92}.  In a cosmological
framework, only waves in a finite range of wavenumbers are
constrained, while longer and shorter waves are unaffected by the
data. In the WF reconstruction this would mean that these waves are
set to zero, while for the constrained realizations these waves are
effectively sampled out of unconstrained realizations. Thus, ideally
the resolution is set to $R_g \approx 2\pi k{_{max}^{-1}}$, where
$k_{max}$ corresponds to the maximum $k$ constrained by the data.

The WF and constrained realizations are constructed assuming that the linear theory is
valid on all scales. Thus, in principle it can be done with any
desired resolution, in particular on scales that at present lie deep
in the non-linear regime. In other words, the WF and constrained realizations provide a
reconstruction or a realization of how the present-day structure would
appear if the linear theory had been valid. This limitation
can be turned into an advantage and used as a tool for recovering the
initial conditions that seeded the growth of structures in the
nearby universe. The basic idea of the present paper is to use the
data on scales where linear theory is applicable to recover the large
scale fluctuations and to supplement them with fluctuations due to a
random realization of a specific power spectrum on small scales.
These fluctuations are extrapolated back in time using the linear
theory to provide a reconstruction of the initial conditions.

\subsection{Observational Data and Prior Model}
\label{subsec:data}

The \m3\ catalog \citep{willick_etal97} has been compiled from several
data sets of spiral and elliptical/S0 galaxies with the direct
Tully-Fisher and the $D_n-\sigma$ distances. The sample consists of
$\approx 3400$ galaxies and provides radial velocities and inferred
distances with fractional errors $\sim 17-21\%$.  The
sampling covers the whole sky outside $\pm (20-30)^\circ$ of the Zone
of Avoidance. It has an anisotropic and non-uniform density that is a
strong function of distance.  The sampling of the density field is
generally good to about $60\hmpc$, although this limit changes from
$40\hmpc$ to $80\hmpc$ for some directions.  The data are
corrected for the Malmquist biases.  A grouping procedure has been
applied to the data in order to lower the inhomogeneous Malmquist bias
before it is used and to avoid strong non-linear effects, in
particular in clusters of galaxies.  This yields a dataset of
distances, radial peculiar velocities, and errors for $\approx 1200$
objects, ranging from individual field galaxies to rich clusters.

The cosmological model assumed here is the currently popular flat
low-density cosmological model ($\Lambda$CDM) with $\Omega_0 = 1 -
\Omega_\Lambda = 0.3$, where $\Omega_0$ is the cosmological density
parameter of the non-relativistic matter and $\Omega_\Lambda$ measures
the cosmological constant $\Lambda$ in units of the critical density.
The Hubble constant is assumed to be $h=0.7$ (measured in units of
$100{\rm\ km s^{-1} Mpc^{-1}}$) and the power spectrum is normalized
by $\sigma_8=0.9$.  The $\Lambda$CDM model assumed here is consistent
with all current observational constraints
\citep[e.g.,][]{wang_etal01}. This model is also consistent with
the radial velocity surveys including the \m3, although the data
favor a slightly higher value of $\Omega_{0}$ \citep{zaroubi_etal97}.

A detailed analysis of the LSS reconstructed from the \m3\ survey was
presented by \citet{zaroubi_etal99}. The most robust features of the
structure recovered from the \m3\ are the Great Attractor (GA), the
Perseus-Pisces (PP) supercluster, the filamentary LSC
connecting GA and PP, and the Local Void. The nearby structures can be
resolved using current methods down to a resolution of about $5\hmpc$.

\subsection{Reconstruction using the \m3\ Survey}
\label{subsec:applic}

\begin{figure}[p!]
\epsscale{2.2}
\plottwo{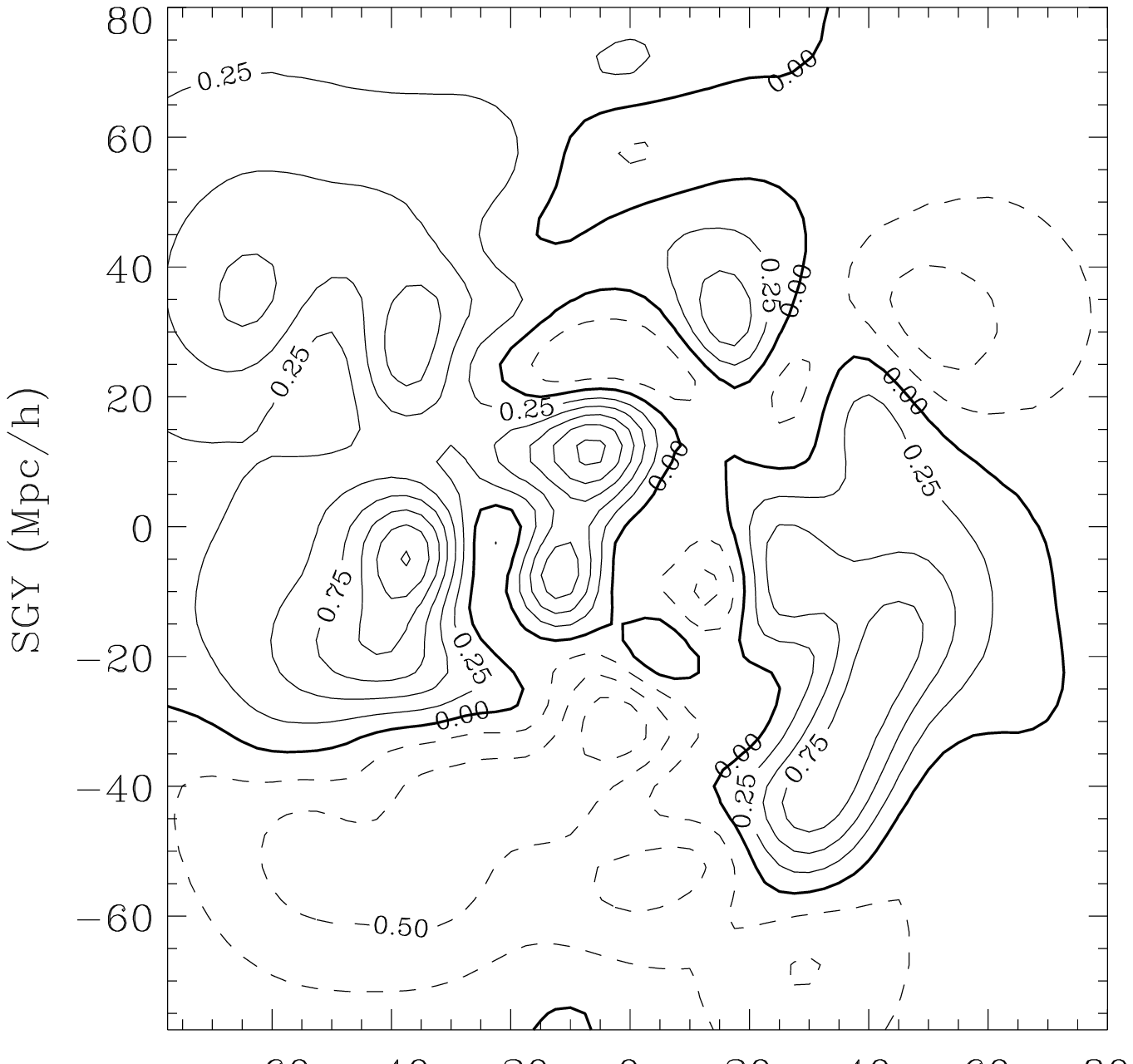}{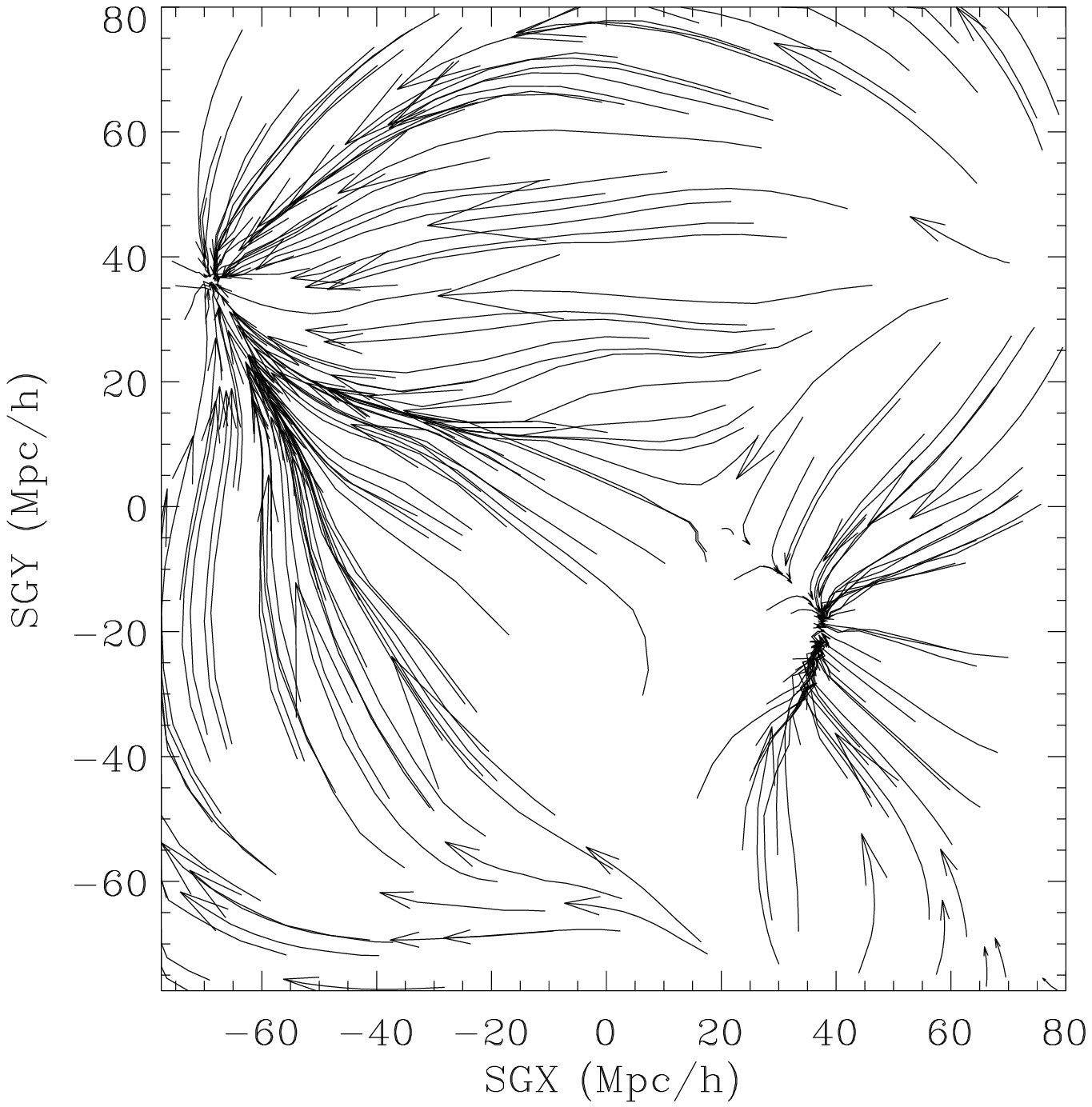}
\caption{\small Wiener filter reconstruction of the density and
  velocity fields from the \m3\ survey of radial velocities, evaluated
  with a $5\hmpc$ Gaussian smoothing. The plots show the structure of
  the supergalactic plane. The density field ({\em top panel}) is
  presented by contours of constant overdensity with spacings of
  $\delta=0.25$ and the velocity field ({\em bottom panel}) is shown
  by streamlines. The LG is in the middle at $[0,0]$. The two
  largest structures are the Perseus-Pisces Supercluster on the low
  right corner and the Great Attractor is in the middle left of the
  plots. }
\label{fig:WF-DV}
\end{figure}

\begin{figure}[p!]
\epsscale{2.2}
\plottwo{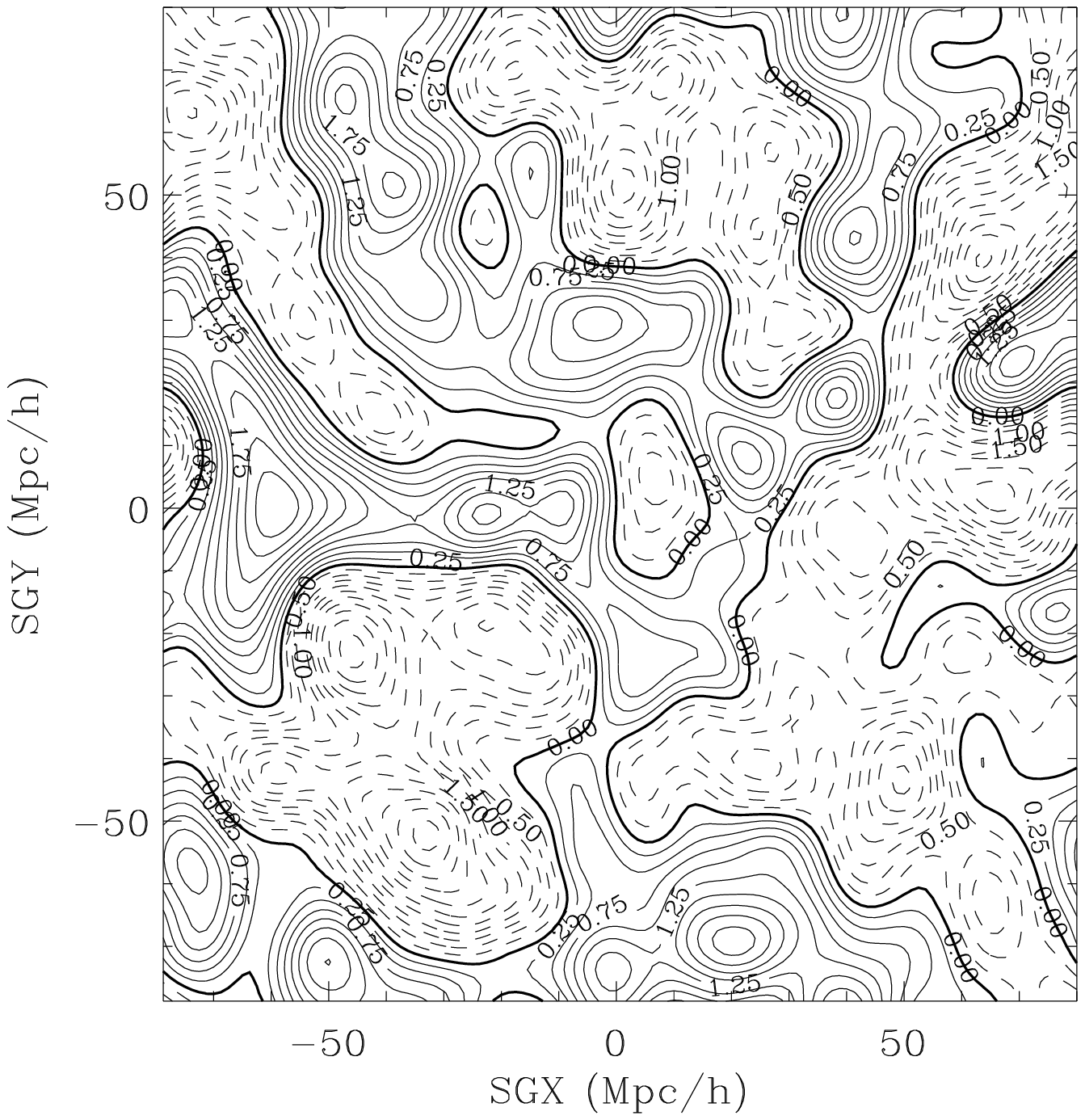}{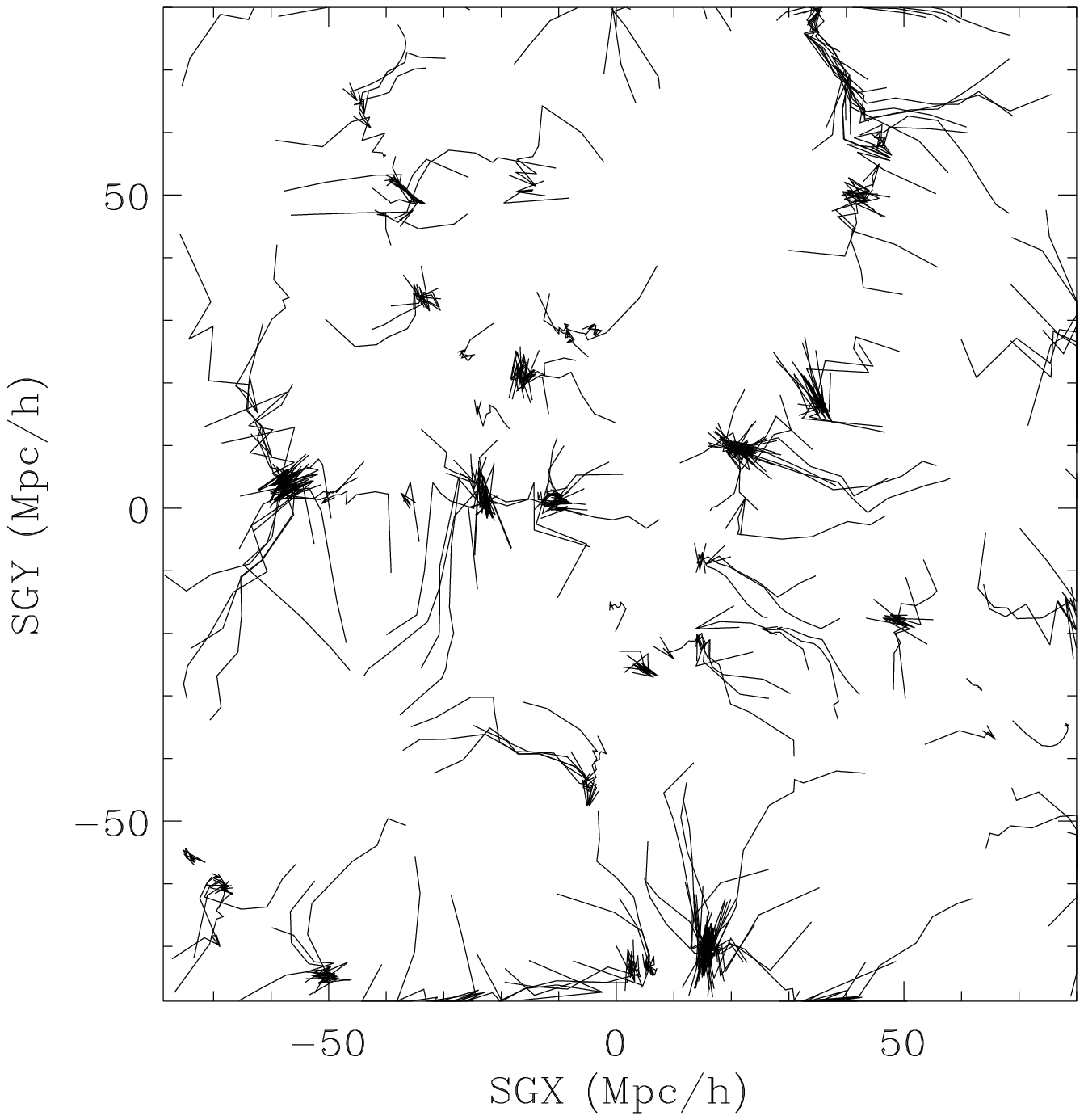}
\caption{\small
  A constrained realization of the density and velocity fields
  constructed using constraints from the \m3\ survey and assuming
  {\LCDM} model. The presentation follows Fig.  \ref{fig:WF-DV},
  except here the velocity field is not smoothed.  In this realization
  the Perseus-Pisces supercluster is located slightly below the
  supergalactic plane (at $SGZ\approx -10\hmpc$) and thus is missing
  in the density plot, but it is apparent in the velocity field. Due
  to periodic boundary conditions the realization is missing a bulk motion of
  $\sim 300~\kms$ in the direction of the Shapley Concentration,
  which is present in the WF velocity field in Figure \ref{fig:WF-DV}.
  The distribution has an overall displacement of about $12\Mpch$ to
  positive $SGY$ with respect to the WF field.  }
\label{fig:CR-DV}
\end{figure}

The WF/Constrained Realizations algorithm has been applied to the \m3\
database assuming a flat $\Lambda$CDM model (see previous
section). This was done on a $128^3$ grid with an $(1.25\hmpc)^3$
cell, thus reconstructing the density and the velocity field within a
box $160\hmpc$ on a side centered on the LG. The WF is evaluated by
calculating the various correlation functions as a one-dimensional $k$
integral over the power spectrum, and therefore the WF is obtained
without imposing any boundary conditions on the grid. On the other
hand, the constrained realizations are based on an unconstrained
realization generated by an FFT algorithm, which imposes periodic
boundary conditions. The density and velocity fields reconstructed
using the WF method are presented in Fig.  \ref{fig:WF-DV}, where the
overdensity, $\delta \equiv \delta\rho / \rho$, and the velocity,
$\bv$, fields are evaluated with the Gaussian smoothing of
$R_g=5\hmpc$. A constrained realization with the same resolution is shown in Figure
\ref{fig:CR-DV}.

Visual comparison of Figures \ref{fig:WF-DV} and \ref{fig:CR-DV}
reveals that there are some similarities and some differences.  On
large scales the velocity fields are more indicative: they show that
the same large structures are present in both distributions. Thus, the
Great Attractor and a large void below it, and the Perseus-Pisces
Supercluster are clearly seen in the velocity field (the PP
supercluster does not appear in the density field because in this
realization the supercluster is outside the shown slice).  The Local
Supercluster is present in both plots as a small ``island'' slightly
above and to the left of the center. There is a small extension
towards negative SGY, which in the real Universe hosts the Fornax poor
cluster and the Eridanus cloud.

Nevertheless, there are some visible differences between Figures
\ref{fig:WF-DV} and \ref{fig:CR-DV}.  The WF field produces a smooth
regular field, by attenuating a large fraction of the power predicted
by the power spectrum. This attenuation increases with the distance
from the LG, corresponding to noisier and sparser data. As a result,
small-scale structures are found only where the data are sufficiently
accurate, as, for instance, in the GA region (around $[SGX, SGY]
\approx[-40,0]\hmpc$. This explains why in the peripheral parts of
Figure \ref{fig:CR-DV} there are many more small structures.  The two
velocity fields differ in one significant aspect: the constrained
realization field clearly reflects the periodic boundary conditions
while WF field does not.  The large-scale bulk velocity is therefore
missing in the case. WF velocity field has a relatively large bulk
flow. This is why the flow around the Perseus cluster in constrained
realization is more symmetric (infall velocities from all directions)
and the flow around the real Perseus cluster is asymmetric: it has net
velocity in the direction of the Great Attractor. Another effect of
the periodic conditions is the shift of the whole matter distribution
in the direction of positive $SGY$ by $11.5\Mpch$ and by $-4\Mpch$ in
$SGX$ direction, compared to the WF field.  This is especially obvious
if one compares the position of the Virgo cluster in WF and in the
constrained realization data. In order to compensate for this, we
displace the whole distribution in supergalactic coordinates by
$\Delta {\bf r}=[3.5,-11.5,0.9]\hmpc$.

One conclusion that follows from this analysis is that the constrained realization adds the
small scale power, while retaining the large scale structures dictated
by the data and manifested by the WF. Generally, the position and
amplitude of objects on scales of a few Mpc do vary with the different
constrained realizations but the large scale ($> 10\hmpc$) structures are reproduced
robustly.

\begin{figure}[t!]
\epsscale{1.001}
\plotone{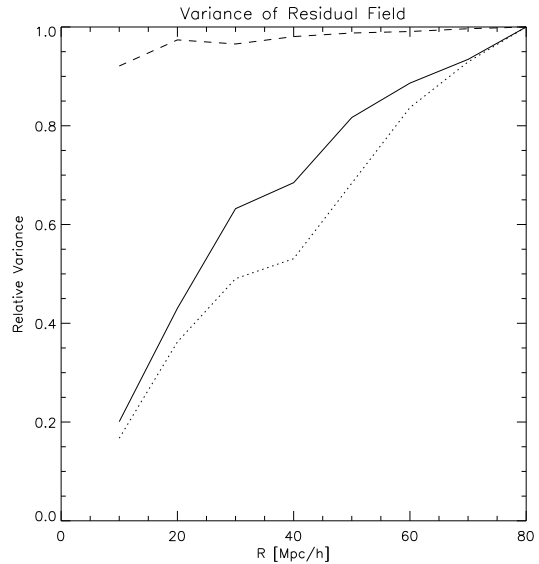}
\caption{\small The mean variance $\langle (\delta_{CR}
  -\delta_{WF})^2\rangle$ of the ensemble of constrained realizations of the density field with respect to the density field obtained using WF reconstruction
as a function of depth for a Gaussian smoothing of
  $R_g=10\hmpc$ (dotted line), $5\hmpc$ (solid line), and no smoothing
  at all (dashed line).  The variance is normalized by the variance of
  the {\LCDM\/} unconstrained field smoothed with the same filter. }
\label{fig:variance}
\end{figure}

To study the constraining power of the data within the framework of
the assumed $\Lambda$CDM power spectrum an ensemble of ten constrained realizations has
been generated. The constrained realizations have been constructed with Gaussian smoothing
of $R_g=5\ {\rm and}\ 10\hmpc$, and with no smoothing at all. We then
estimate the mean variance $\langle (\delta_{CR}
-\delta_{WF})^2\rangle$ of the constrained realizations (the scatter around the WF $\delta$
field) for all regions inside a sphere of radius $R$ from the LG.  By
comparing the smoothed density field of a constrained realization to the original WF field,
we can gauge the power of the observational constraints in the 
realization.  Figure \ref{fig:variance} shows the variance normalized
by the unconstrained expected variance 
$\left<\delta^2_{\Lambda {\rm CDM}}\right>$.
The plot shows that down to a resolution of a few Mpc the 
region within $\approx 20\hmpc$, which is the subject of the present
study, is well constrained by the data: the variance is small with
only 20--25\% of the power contributed by the random component of the
constrained realizations. This ratio increases with the depth; at large distances the
realizations become unconstrained. On the other hand, on the scales of
$\lesssim 5-10\hmpc$, the fluctuations are heavily dominated by the
random realization and are virtually unaffected by the constraints
(the field is unconstrained on scales below the grid cell, $\lesssim
1\hmpc$). It follows that the unsmoothed constrained realization provides us with a
realization that has all the power demanded by the assumed model,
with the large scales being constrained by the data, while the small-scale
waves are largely unconstrained.

\subsection{Constrained Simulations}
\label{subsec:CS}

The aim of this paper is to perform $N$-body simulations that match the
observed local universe as well as possible. Namely, we are interested
in reproducing the observed structures: the Virgo cluster, the Local
Supercluster and the LG, in the approximately correct
locations and embedded within the observed large-scale configuration
dominated by the GA and PP.  constrained realizations, however, show some sizeable
deviations from the WF-reconstructed field and the degree with which
a given realization matches the observed density and velocity fields varies. These
deviations are, of course, much smaller than in the case of an
unconstrained realization, but we would nevertheless like to
concentrate the computational effort on a realization that matches the
observed configuration most closely.  

To this end, we have generated five constrained realizations and
selected a realization (shown in Fig.~\ref{fig:CR-DV}) that provides
the best match to the observed density field reconstruction.
Specifically, we have performed low-resolution simulations of each
constrained realization.  These simulations were run with $128^3$
equal mass particles with a particle mass of $1.6\times 10^{11}\Msunh$
and the highest formal spatial resolution of $20\kpch$.  All the
simulations have reproduced the Virgo cluster, the Great Attractor and
the Perseus-Pices supercluster, with the Virgo cluster being embedded
in a LSC.  The position of the Virgo cluster varied in the simulations
along the LSC by more than $10\hmpc$ either in the direction of the GA
or to the PP supercluster. We have selected the realization in which
the Virgo cluster at the present epoch was in an approximately correct
location.  This was the only selection criterion as the resolution of
the simulations was barely sufficient to resolve the Virgo
cluster. The small-scale structure within the LSC region was not
resolved.

Using the selected constrained realization, two simulations with increasing force and mass
resolution in the region around the Virgo Cluster were performed.  The
initial conditions for these simulations were set using multiple mass
resolution. At the $z=0$ output of the low-resolution run, we selected all
particles within a sphere of $25\hmpc$ radius centered on the Virgo
cluster.  The mass resolution in the lagrangian region occupied by the
selected particles was increased and additional small-scale waves
from the initial {\LCDM} power spectrum of perturbations were added
appropriately \citep[see ][ for details of the method]{kkbp01}. For
the two high-resolution simulations, the particle mass in the LSC
region is 8 and 64 times smaller than in the low-resolution
simulation.  The highest resolution simulation has particle mass
$2.51\times 10^9\Msunh$ and the maximum formal force resolution was
$2.44\kpch$ in the LSC region.  The results of both high-resolution
simulations agree well with each other at all resolved scales. Below
we present results only from the highest resolution run.

\section{Numerical simulations}
\label{sec:numer}

The ART $N$-body code \citep{kkk97,kravtsov99} was used to run the
numerical simulation analyzed in this paper.  The code starts with a
uniform grid, which covers the whole computational box. This grid
defines the lowest (zeroth) level of resolution of the simulation.
The standard Particles-Mesh algorithms are used to compute the density and
gravitational potential on the zeroth-level mesh.  The code then
reaches high force resolution by refining all high density regions
using an automated refinement algorithm.  The refinements are
recursive: the refined regions can also be refined, each subsequent
refinement having half of the previous level's cell size. This creates
a hierarchy of refinement meshes of different resolution, size, and
geometry covering regions of interest. Because each individual cubic
cell can be refined, the shape of the refinement mesh can be arbitrary
and effectively match the geometry of the region of interest. This
algorithm is well suited for simulations of a selected region within a
large computational box, as in the constrained simulations presented
below.

The criterion for refinement is the local density of particles: if the
number of particles in a mesh cell (as  estimated by the Cloud-In-Cell
method)  exceeds  the   level  $n_{\rm thresh}$,  the   cell  is split
(``refined'')  into  8   cells of   the   next refinement  level.  The
refinement threshold may depend on the refinement level. The code uses
the   expansion  parameter  $a$ as  the    time variable.   During the
integration,   spatial  refinement is    accompanied  by  temporal
refinement.  Namely, each level of refinement, $l$, is integrated with
its own time  step $\Delta a_l=\Delta  a_0/2^l$, where $\Delta a_0$ is
the global time step  of the zeroth  refinement level.   This variable
time stepping  is very important for  accuracy of the results.  As the
force resolution  increases, more  steps  are needed to integrate  the
trajectories accurately.  Extensive  tests of the code and comparisons
with other  numerical $N$-body codes can  be found  in 
\citet{kravtsov99} and \citet{knebe_etal00}.

The current version of the ART code has the ability to handle
particles of different masses.  In the present analysis this ability
was used to increase the mass (and correspondingly the force)
resolution inside a region centered around the Virgo cluster.  The
multiple mass resolution is implemented in the following way.  We
first set up a realization of the initial spectrum of perturbations in
such a way that initial conditions for a large number ($1024^3$) of
particles can be generated in the simulation box.  Coordinates and
velocities of all the particles are then calculated using all waves
ranging from the fundamental mode $k=2\pi/L$ to the Nyquist frequency
$k=2\pi/L\times N^{1/3}/2$, where $L$ is the box size and $N$ is the
number of particles in the simulation.  Some of the particles are then
merged into particles of larger mass and this process can be repeated 
for merged particles. The larger mass (merged)
particle is assigned a velocity and displacement equal to the average
velocity and displacement of the smaller-mass particles.

The simulations presented here were run using $256^3$ zeroth-level
grid in a computational box of $160\Mpch$.  The threshold for cell
refinement (see above) was low on the zeroth level: $n_{\rm
  thresh}(0)=2$.  Thus, every zeroth-level cell containing two or more
particles was refined.  This was done to preserve all small-scale
perturbations present in the initial spectrum of perturbations.  The
threshold was higher on deeper levels of refinement: $n_{\rm
  thresh}=3$ and $n_{\rm thresh}=4$ for the first level and higher
levels, respectively.

For the low resolution runs the step in the expansion parameter was
chosen to be $\Delta a_0=3\times 10^{-3}$ on the zeroth level of
resolution.  It was $\Delta a_0=2\times 10^{-3}$ for the high
resolution runs.  This gives about 500 steps for particles located in
the zeroth level for an entire run to $z=0$ and 128,000 for particles
at the highest level of resolution.

\section{Results}
\label{sec:results}

\subsection{Morphology of matter distribution}
\label{subsec:matter}

\begin{figure*}[p]
\includegraphics[width=\textwidth]{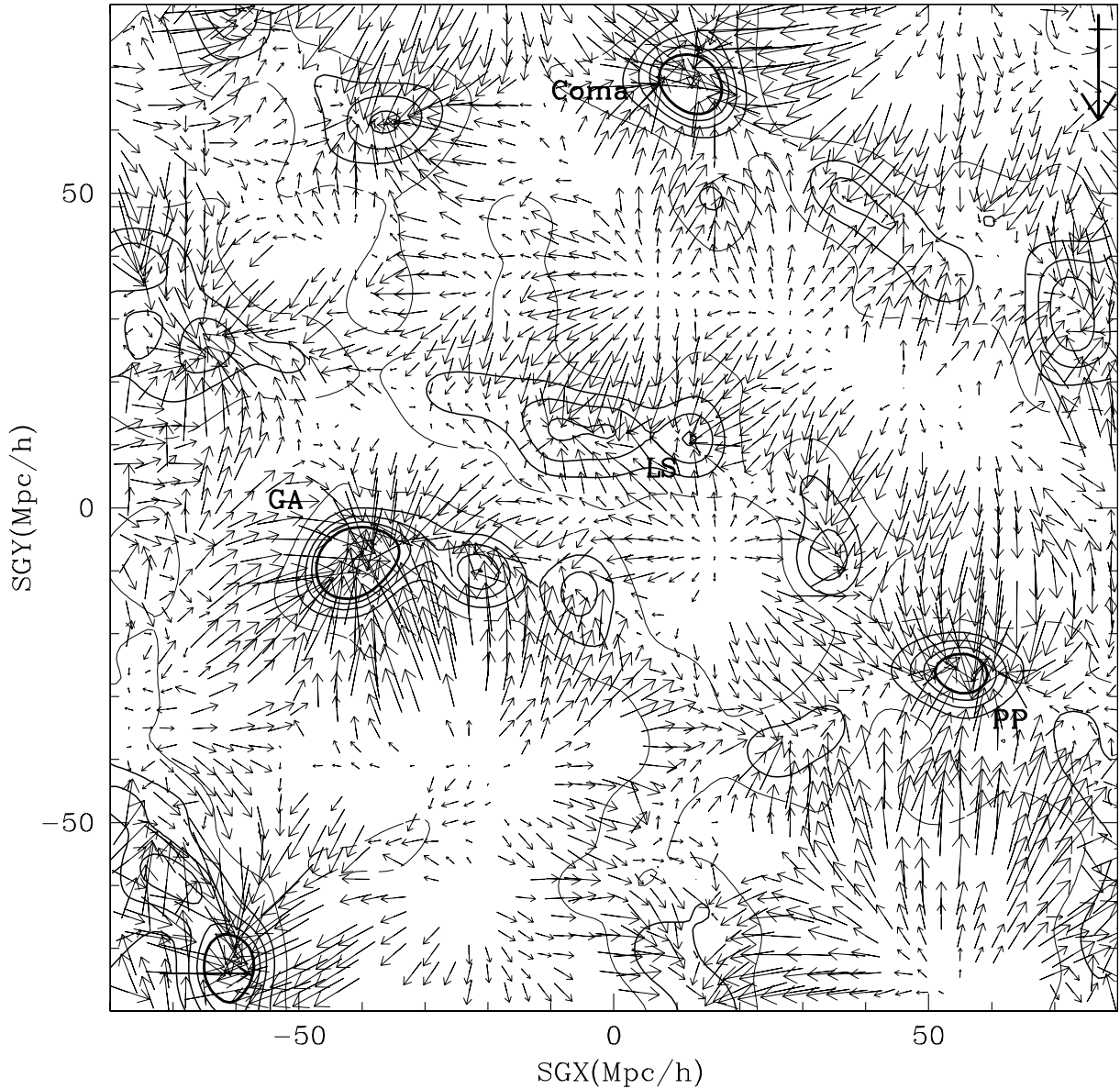}
 \caption{\small Density and velocity fields in a slice 
   through the central regions of the simulated volume ($SGZ=0$). The
   fields are smoothed with the $5\hmpc$ Gaussian filter. Positions of
   several known objects are marked (GA - Great Attractor, LS - Local
   Supercluster, PP - Perseus-Pisces Supercluster, Coma - Coma
   cluster).  The thick large arrow in the top
   right corner corresponds to the velocity of $1000~\kms$. The thin
   density contours correspond to 1, 2, 3, and 4 times of the average
   density of the matter in the Universe, while the thick contours
   shows the overdensity 5. Our position is at the origin of the
   system of coordinates at (0,0).  Due to a rather large smoothing
   length, most of the structures prominent in this figure are well
   constrained to match the observed structures in the Mark III
   catalog.  An important difference between the simulated and
   observed velocity fields is that the tidal field produced by the
   matter outside the simulated volume was removed.
   }\label{fig:SliceAll}
\end{figure*}

Figure~\ref{fig:SliceAll} shows the density and velocity fields in a
slice centered on $SGZ=0$.  The fields are smoothed with the Gaussian
filter of $5\hmpc$ smoothing length.  
No additional projection was
used. The plane of the slice was rotated by 30 degrees around the
vertical line $SGX=0$. This slice goes through both the Great
Attractor and the Perseus-Pisces supercluster. 
 All major structures
(the LSC, Great Attractor, Perseus-Pisces supercluster, and Coma
cluster) observed within $100\hmpc$ around the Milky Way exist in the
simulations. The positions and morphology of these structures is, of
course, fairly well constrained by the constraints imposed on the
initial conditions.  In this projection the LSC is an
elongated structure some $15\hmpc$ above the origin. It extends over
$\sim 40\hmpc$ along the SGX axis. There is a low-density ``bridge''
(of overdensity just above the average density), which connects the
LSC with the Perseus-Pisces Supercluster. There is a also an even
weaker filament connecting the LSC with the Great
Attractor (it is not obvious in Fig.~\ref{fig:SliceAll} but is visible
on the left side of Fig.~\ref{fig:SliceLS}).

Figure~\ref{fig:SliceLS} shows a zoom-in view of $45\hmpc$ region
around the Virgo cluster. It is clear that the matter distribution is
very far from being spherically symmetric. The main body of the Local
Supercluster is a filament roughly centered on the Virgo Cluster and
extending from the Ursa Major cluster to a concentration of several
massive groups (region around (-10,10)\Mpch). Smaller filaments
connect the LSC to other nearby structures.  Two filaments in the
upper half of the plot (SGX$\approx -16\Mpch$ and $\approx 14\Mpch$)
connect the LSC with the Great Wall and the Coma cluster. The filament
extending diagonally from the Ursa Major down to the bottom right
corner of the slice forms a ``bridge'' to the Perseus-Pices
Supercluster. The weak filament extending almost straight down along
the SGY axis from the Virgo cluster connects it to the LG and,
eventually, to the Great Attractor. The large structure at the bottom
of the figure at $(SGX, SGY) =(-8, -10)\Mpch$ is the location of the
Fornax cluster counterpart in the simulation.  It was just outside of
the region of high mass and force resolution. The sharp decline in the
number density of particles in the bottom left and right corners of
the Figure~\ref{fig:SliceLS} marks the boundary of the high resolution
region.

 Just as in the real
Universe, the LG is located in a weak filament extending
between the Virgo and Fornax clusters. This filament is a counterpart
of the Coma-Sculptor ``cloud'' in the distribution of nearby galaxies
\citep[see][for the survey of nearby structures]{tully_fisher87}.
Figure~\ref{fig:SliceLS} shows that our Galaxy is not located in a
common galaxy environment. A typical galaxy is very likely to be found
in the main body of the LSC few megaparsecs away from a large group or
a cluster. At the same time, our location is not very special. There
are other small filaments in LSC, which are similar to our
Coma-Sculptor ``cloud''.

Figure~\ref{fig:SliceLG} shows a zoom-in view of the immediate
environment of the simulated LG. Note that the structures at
these scales are only weakly affected by constraints imposed on the
initial conditions. Several possible counterparts to existing objects
(e.g., the MW and M31, M51, NGC253) are marked, but their existence is
largely fortuitous. As can be seen in this figure, the simulated LG is
located in a rather weak filament extending to the Virgo cluster (see
Fig.~\ref{fig:SliceLS}). This filament borders an underdense region
visible in the right lower corner of Figure~\ref{fig:SliceLG}, which
corresponds to the Local Void in the observed distribution of nearby
galaxies. Note that the velocity field around the LG is
rather quiet. The peculiar velocity field in the Local Void exhibits a
uniform expansion of matter out of this underdense region, while
velocities between the LG and the Virgo (upper half of
Fig.~\ref{fig:SliceLG}) show a coherent flow onto the main body of the
LSC.

As can be seen from Figures~\ref{fig:SliceAll}-\ref{fig:SliceLG} most
of the mass in the LSC region is concentrated in a filament centered on the
Virgo cluster, while the LG is located in an inconspicuous
albeit slightly overdense region neighboring the Local Void. The
matter distribution is thus not spherically symmetric as is often
assumed in the Virgocentric velocity flow models and the mass within
the Virgo cluster itself constitutes only a small fraction of the
LSC's mass.  In the next section we will discuss the properties of the
peculiar velocity field in this region and compare it to the observed
velocity of nearby galaxies.

\subsection{Peculiar velocity field in the LSC region}
\label{subsec:vel}

\begin{figure*}[p]
\includegraphics[width=\textwidth]{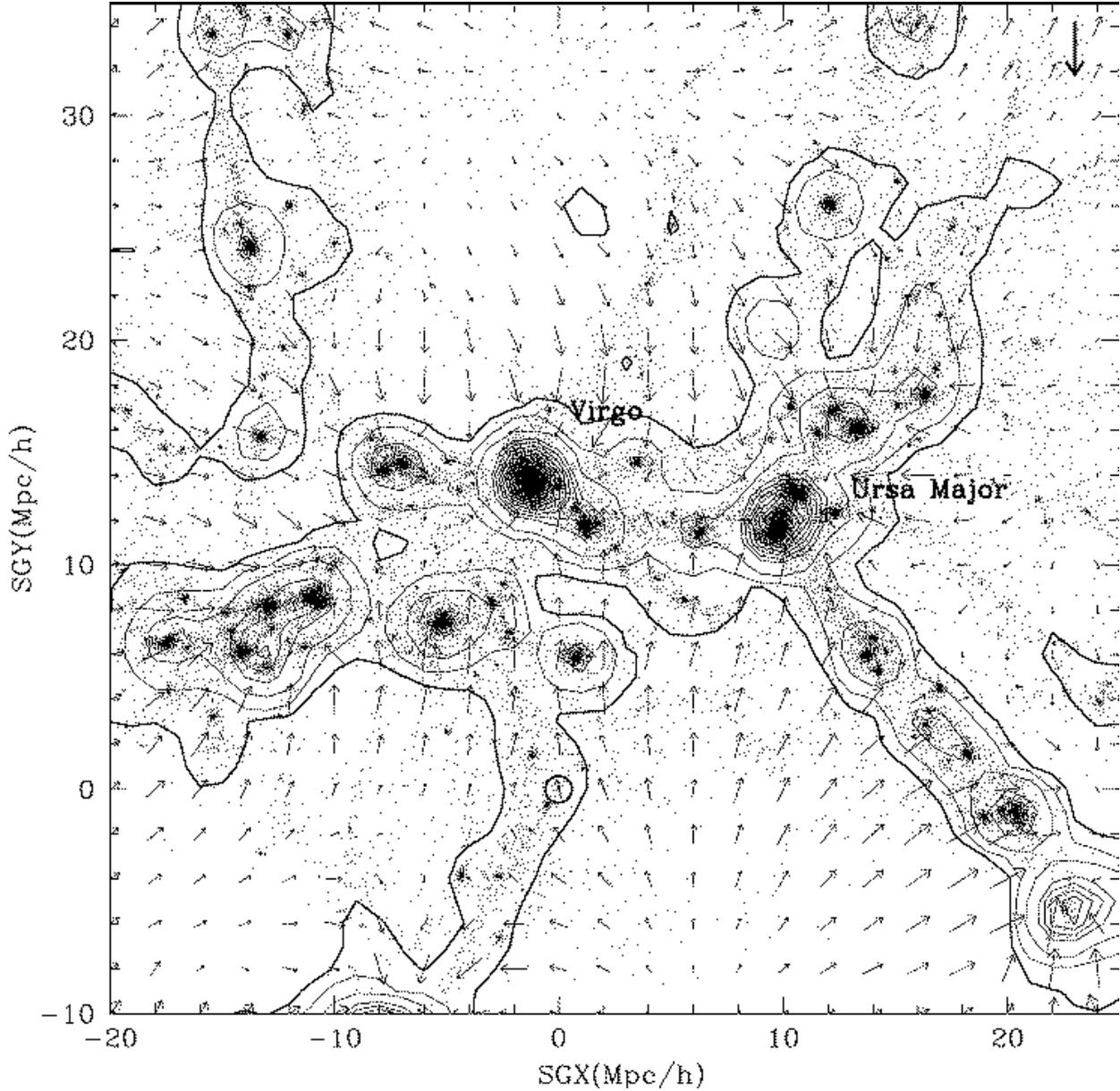}
\caption{\small Density and velocity fields in the 45\Mpch~ region around the
  Virgo cluster. The velocities are in the Virgo cluster rest frame.
  The fields were smoothed with the Gaussian filter of 1.4\Mpch~
  smoothing length. The circle at the origin of the coordinates marks
  the position of the Milky Way galaxy. Points show dark matter
  particles (10\% of all particles is shown) in a slice of $10\Mpch$
  thickness centered on the Supergalactic plane ($SGZ=0$). Contours
  show the projected density in the slice: the thick contour corresponds
  to the average matter density of the Universe. The thin contours
  mark overdensities 2, 4, 6, and so on.  The length of the thick
  arrow in the top right corner shows a velocity of $500~\kms$.
  }\label{fig:SliceLS} 
\end{figure*}

\begin{figure*}[p]
\includegraphics[width=\textwidth]{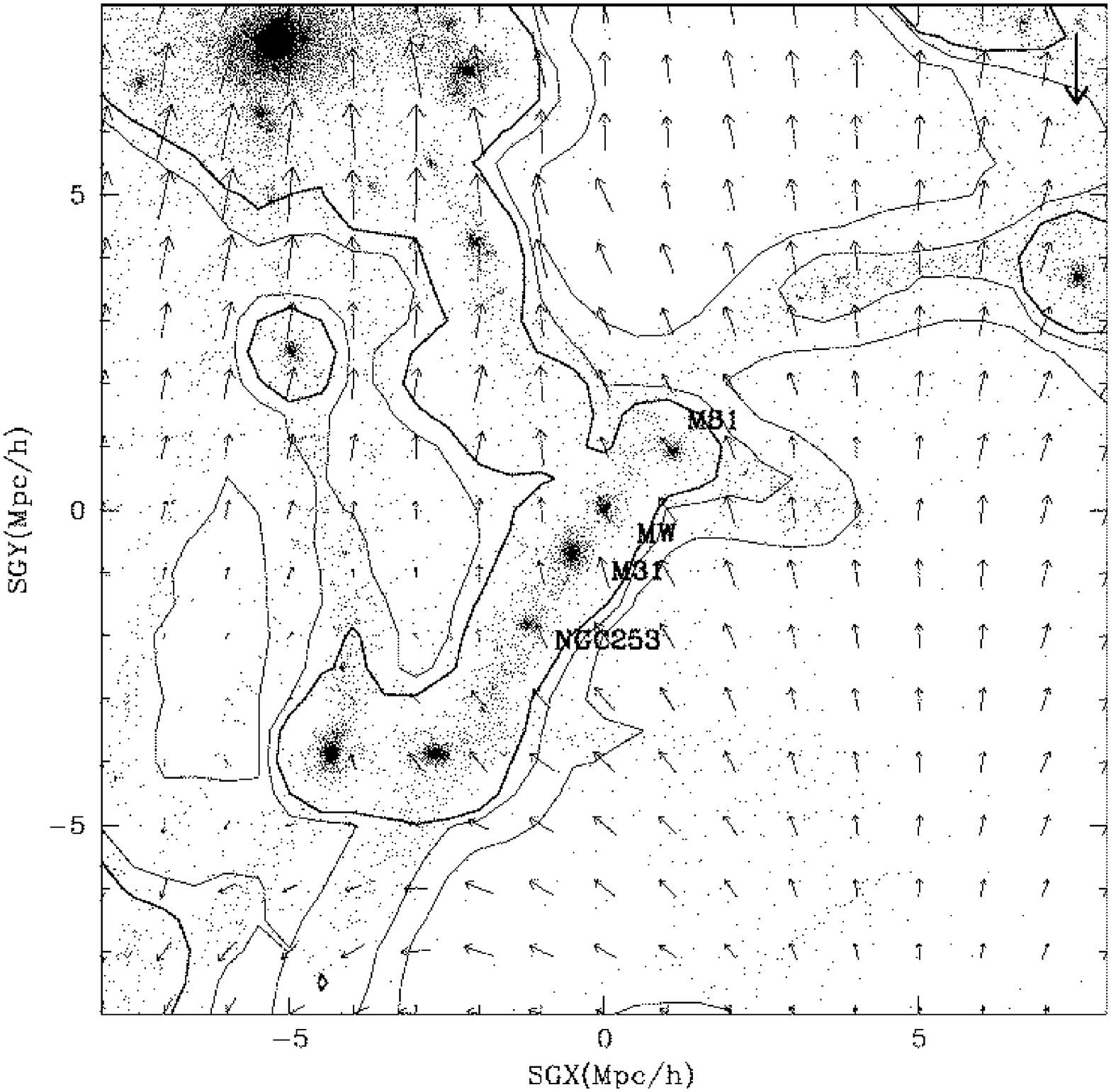}
\caption{\small Density (contours) and velocity (arrows) fields (smoothed with a Gaussian filter of $0.7\hmpc$ smoothing length) around the LG. 
  The slice shown  has a size and thickness of $15\hmpc$ and $5\hmpc$,
  respectively, and is centered on the supergalactic plane ($SGZ=0$).
  Points show positions of the dark matter particles (all particles are
  shown). Contours show the projected density in the slice: the thick
  contour corresponds to the overdensity 3; the other contours mark
  overdensities 1 and 2. The length of the thick arrow in the top
  right corner corresponds to a velocity of $500~\kms$. The velocities are
  plotted in the Virgo cluster rest frame. Although the matter
  distribution bears some resemblance to the observed distribution of
  galaxies, observational data constraints on these scales are weak.
  Note the empty region to the right of the LG. 
  }\label{fig:SliceLG}
\end{figure*}

Figures~\ref{fig:SliceAll}-\ref{fig:SliceLG} show that the peculiar
velocity field in the LSC region is quite complex. The
deviations from the uniform Hubble flow can be thought of as two
different components: the bulk flows (coherent large-scale flows which
can be seen in Figs.~\ref{fig:SliceAll}-\ref{fig:SliceLG}) and the
small-scale velocities in and around collapsed (or collapsing)
objects.  In underdense regions the bulk velocities exhibit a roughly
spherically symmetrical pattern typical of an expanding void
\citep{bert85}: the peculiar velocities are steadily increasing from
the center of the void to its bordering filaments.  In
Figure~\ref{fig:SliceLS} this pattern can be seen clearly in the two
voids in the top (between the Virgo and the Coma clusters) and the
lower right corner (the Local Void). Note that in the vicinity of the
high-density regions peculiar velocities tend to have direction
perpendicular to the nearest pancake or filament (see, for example,
velocity field around the filament near the ``Ursa Major cluster'' in
the middle right of Fig.~\ref{fig:SliceLS}). This behavior can be well
understood in terms of the \citet{zeldovich70} pancake solution which
predicts velocity flow in the direction perpendicular to the largest
dimension of the collapsing pancake.

It is also clear that peculiar velocities do not exhibit a simple
Virgocentric spherically symmetric infall pattern usually assumed in
the models of the local velocity field
\citep[e.g.,][]{aaronson_etal82,huchra88}.  Only in the immediate
vicinity of the Virgo cluster (within $\sim 3$~Mpc of the cluster
center) the velocity infall is Virgocentric and is close to being
spherically symmetric. On larger scales the velocity flow is 
roughly perpendicular to the filament which constitutes the main 
body of the LSC, reflecting the fact that most of the mass is distributed
in this filament rather than being concentrated in one cluster.

\begin{figure}[tb!]
\epsscale{0.95}
\plotone{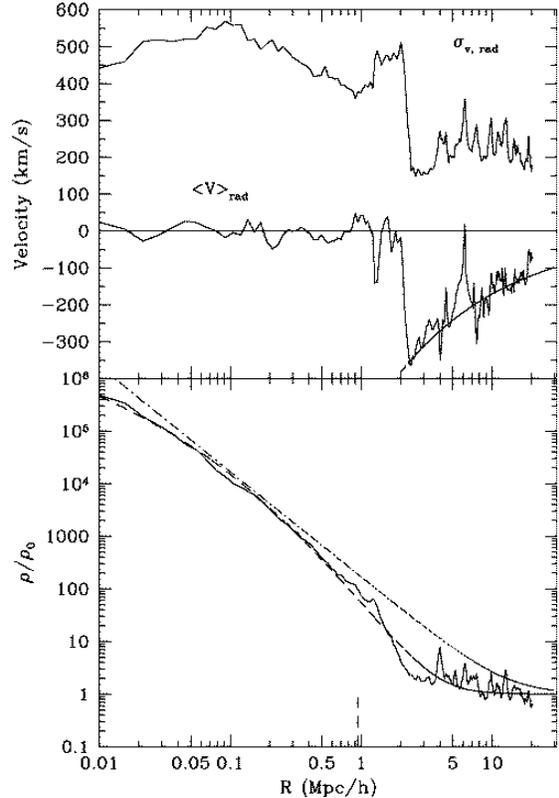}
\caption{\small Spherically averaged profiles of density ({\em bottom
panel}), average radial velocity ({\em top panel, lower curve}), and
radial velocity dispersion ({\em top panel, upper curve}) constructed
around the center of the Virgo cluster.  {\em Bottom panel:} The
density profile is shown in units of the average density of the
universe, $\rho_0$. The short vertical dashed line on the x-axis at
$\sim 1\Mpch$ indicates the virial radius of the Virgo Cluster. The
dashed curve shows the NFW profile \citep{NFW97} with an added
constant $\rho_0$ to account for the flattening of the profile at
large radii. The dot-dashed curve shows the density model assumed by
Huchra (1988) to model the velocity field in the LSC region.  {\it Top
panel:} note that the average radial velocity of the dark matter is
very small within the virial radius of the Virgo cluster, which
indicates an established virial equilibrium within the cluster. The
dashed curve outside the virial radius shows a fit to the infall
velocity profile outside the Virgo cluster:
$145(13\Mpch/r)^{1/2}~\kms$.  }\label{fig:VirgoProfile}
\end{figure}

The predictions of the Virgocentric infall model fail to describe the
properties of the simulated peculiar velocity field accurately both
qualitatively and quantitatively.  Figure~\ref{fig:VirgoProfile} shows
spherically averaged density, radial velocity, and velocity dispersion
profiles centered on the Virgo cluster\footnote{The simulated Virgo
cluster has the virial radius $960\kpch$ indicated by short vertical
dashed line in Figure~\ref{fig:VirgoProfile}, the virial mass of
$M_{\rm vir}=1.03\times 10^{14}\Msunh$, and the concentration of
$C=8.7$ \citep[see ][ for definition]{NFW97}.}. The average density
contrast at the distance of the LG in the simulation is
$\delta\rho/\rho=0.69$. The dot-dashed line in
Figure~\ref{fig:VirgoProfile} shows the density model used by
\citet{huchra88} in the Virgocentric infall model: a constant density
with added $r^{-2}$ law, which is normalized to have an overdensity of
4 at the LG distance.  This model overpredicts the mass within the MW
distance around the Virgo cluster in our simulation by a factor of
three.  This is consistent with the results of
\citet{villumsen_davis86}, \citet{cen94}, and \citet{governato_etal97}
who concluded that the Virgocentric infall model is quite inaccurate
when tested using numerical simulations.

Figure~\ref{fig:VirgoProfile} also shows that the average radial
velocity of the dark matter is near zero within the virial radius of
the Virgo cluster, which indicates that cluster matter is in virial
equilibrium. The deviations of the average velocity from zero at
larger radii shows that the matter near and just outside the virial
radius is not fully relaxed. At even larger radii the profile shows
the infall of unvirialized matter and can be well approximated as
$145(13\Mpch/r)^{1/2}~\kms$ (shown by the dashed curve).  The radial
velocity dispersion profile in the top panel of
Figure~\ref{fig:VirgoProfile} is approximately flat at large radii
(indicating the velocity dispersion typical in the LSC region), while
within the virial radius the velocity dispersion profile has a shape
typical for the halos formed in the CDM cosmologies
\citep{NFW97,lokas_mamon01,kkgk99}.

\begin{figure}[tb!]
\epsscale{1.04}
\plotone{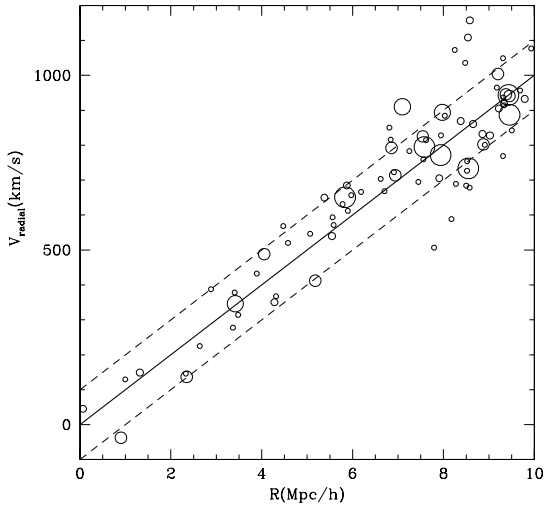}
\caption{\small The Hubble diagram (radial velocity as a function of radius) 
  for halos in the simulation in the LG restframe.  Individual
  halos are denoted by open circles with the circle radius
  representing the maximum circular velocity of the halo (larger
  circles correspond to more massive halos). The thick line shows the
  Hubble law, $v_{\rm H}=H_0 r$, for the value $H_0=70{\ \rm km\ s^{-1}
    Mpc^{-1}}$ assumed in the simulation, and the thin lines
    correspond to velocities $v_{\rm H}\pm 100{\ \rm km\ s^{-1}}$.
    Note that deviations from the Hubble flow in the vicinity of the
    LG ($r\lesssim 7$\Mpch) are $<100{\ \rm km\ s^{-1}}$ 
   (i.e., the flow is ``cold''). }
  \label{fig:LocalFlow}
\end{figure}

Figure~\ref{fig:LocalFlow} shows the Hubble diagram for the simulated
DM halos in the Milky Way restframe. Similarly to observations
\citep[e.g.,][]{sandage99}, galactic halos follow the global Hubble
expansion remarkably well. Note that deviations from the Hubble flow
in the vicinity of the LG ($r\lesssim 7$\Mpch) are
$<100~\kms$ and the flow is rather ``cold''. In particular, similarly
to the observed local Hubble flow
\citep[e.g.,][]{giraud86,ekholm_etal01} only one halo (besides the M31
counterpart) has negative radial velocity (i.e., is blueshifted).
Peculiar velocities increase at larger ($r\gtrsim 7$\Mpch) radii due
to the large concentration of mass in the main body of the LSC, in the
immediate vicinity of the Virgo cluster. Table~\ref{tab:sigmar}
presents the radial and 3-dimensional velocity dispersions in
different radial shells around the MW for halo samples constructed
using various selection criteria.  The columns show the shell radius
(1), radial velocity dispersion (2), 3-dimensional velocity dispersion
(3), number of halos in the radial shell, halo sample selection
criteria (5, with $N_p$ and $V_{\rm circ}$ denoting the number of
particles per halo and the circular velocity of the halo,
respectively). The rms radial velocity at $r=5\hmpc$ is about
$60~\kms$ (virtually independent of the sample selection criteria) and
increases to $\approx 110~\kms$ at $r=10\hmpc$.

\begin{figure}[tb!]
\epsscale{0.99}
\plotone{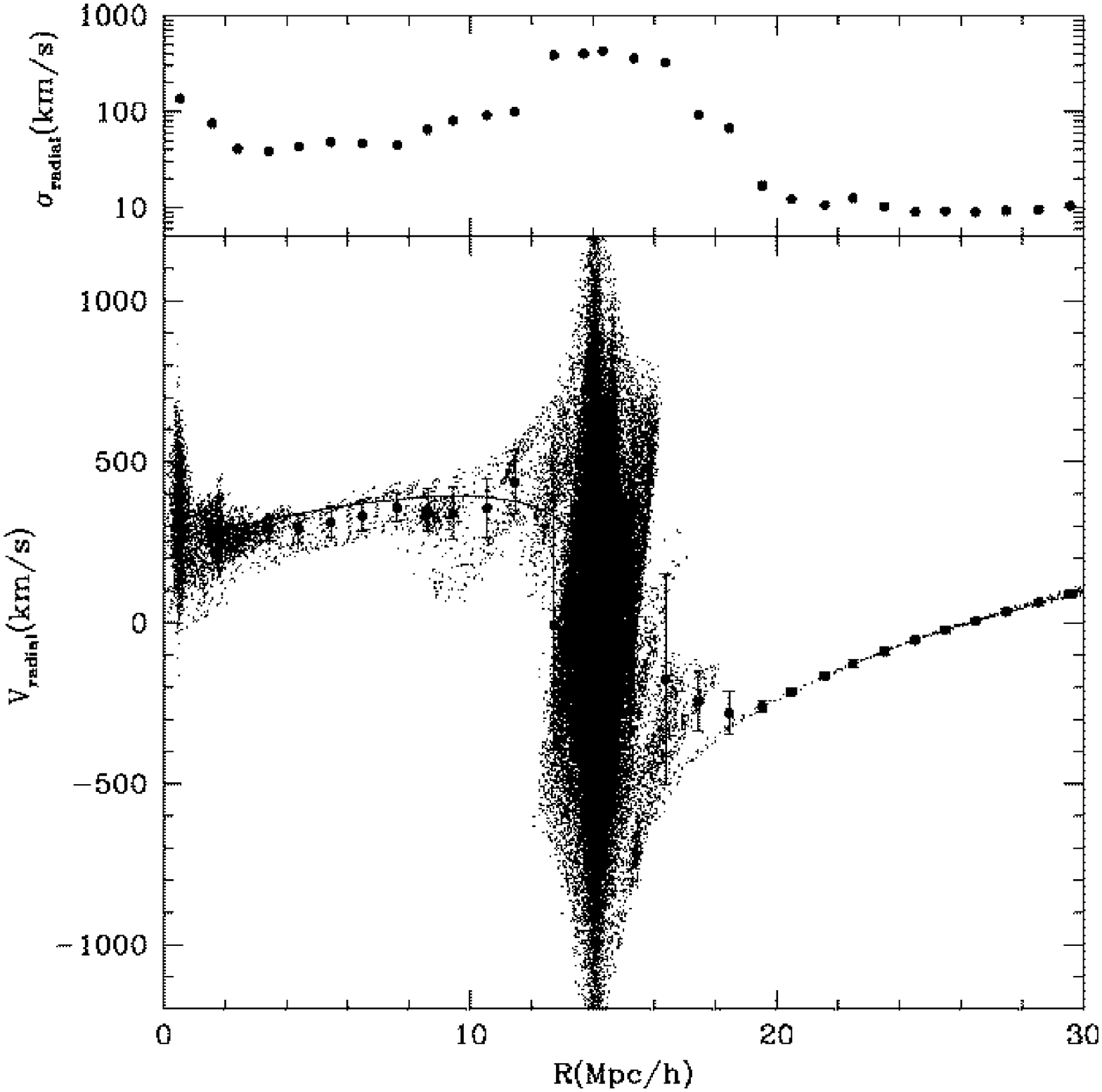}
\caption{\small {\em Bottom panel:\/} Peculiar radial velocities of DM
particles (points) along the line-of-sight from the LG through the
center of the Virgo cluster. All particles within the radius of
$1.5\hmpc$ of the line-of-sight are plotted.  Solid circles with error
bars show average and the rms deviation of the radial velocities as a
function of radius; the rms radial velocity dispersion is shown
separately in the {\em top panel}.  The Virgo Cluster is at the
distance of $14\Mpch$ from the LG.  The particle velocities
are computed in the restframe of the Virgo cluster.  }
\label{fig:LocalVoid}
\end{figure}

To illustrate deviations from spherical symmetry in the velocity flow
around the Virgo cluster, Figure~\ref{fig:LocalVoid} shows the peculiar
radial velocities of DM particles along the line-of-sight passing
through the the Virgo Cluster. All DM particles within $1.5\hmpc$
around the line-of-sight are shown and particle velocities are
computed in the restframe of the Virgo cluster. The most prominent
feature in the radial velocity profile is the ``finger of god'' effect
within the LG and the Virgo cluster. This well-known effect is due to
large random velocities of particles within these massive halos. The
LG and the Virgo cluster are the only massive systems along this
line-of-sight.  Therefore, the radial velocity dispersion of particles
outside the virialized regions measure the ``temperature'' of the 
intergalactic velocity field. Note that the typical deviations from
the coherent infall are very small on both sides of the Virgo Cluster:
$\sim 50-60~\kms$ at $r\lesssim 7\hmpc$ and $\lesssim
10~\kms$ at $r\gtrsim 18\hmpc$, which reflects the
relative ``coldness'' of the flow. The flow is especially cold at
large radii because the matter at these radii undergoes a fairly uniform
expansion out of a void.

The solid curve in Figure~\ref{fig:LocalVoid} represents a model for
the infall velocity based on the Zeldovich approximation. The model
was normalized to have the same infall velocity at the distance from
the LG to the center of the Virgo cluster as found in the
simulation. This model fares much better than the spherical
Virgocentric infall model of \citet{huchra88}: it makes good fit for
all velocities at all radii between the LG and the
Virgo. Nevertheless, it failed because it required a mass between the
LG and the Virgo that is about twice smaller than the corresponding
mass in the simulation (compared to the factor of three overprediction
of mass by the Virgocentric infall model).  Moreover, the velocity
flow on the other side of the Virgo cluster at $r\gtrsim 15\hmpc$ is
not reproduced by this model even qualitatively.  In fact, the
velocity flow around the Virgo cluster is clearly asymmetric and
cannot be well described by a model with any kind of symmetry with
respect to the Virgo cluster. This effort illustrates how difficult it
is to construct a reasonable model for the velocity flow in the Local
Supercluster.

\begin{deluxetable}{ccccl}
\tablecolumns{5}
\tablewidth{0pc}
\tablecaption{Peculiar velocity dispersions in the vicinity
of the LG}
\tablehead{
\colhead{\small Distance from the LG} & \colhead{$V_{\rm radial}$} &
\colhead{$V_{\rm 3D}$}& \colhead{$N_{\rm halos}$} &
\colhead{conditions}  \\
\colhead{\Mpch}   & \colhead{$\kms$}    & \colhead{$\kms$} &  & \\
}
\startdata
~5 & $60\pm 16$ & 129 & 14 & $N_p>10$ \& $V_{\rm circ} > 55~\kms$ \\
~5 & $58\pm 16$ & 130 & 13 & $N_p>15$ \hfill \\
~5 & $57\pm 16$ & 132 & 12 & $N_p>20$ \& $V_{\rm circ} > 60~\kms$ \\
~5 & $48\pm 20$ & 136 & 6 & $N_p>25$ \& $V_{\rm circ} > 90~\kms$ \\
~6 & $66\pm 14$ & 139 & 24 & $N_p>15$\hfill \\
~6 & $67\pm 14$ & 141 & 17 & $N_p>25$ \hfill\\
~8 & $84\pm 13$ & 165 & 43 & $N_p>15$ \hfill\\
~8 & $77\pm 14$ & 181 & 29 & $N_p>25$\hfill \\
10 & $111\pm 13$ & 222 & 75 & $N_p>15$ \hfill\\
10 & $110\pm 15$ & 222 & 55 & $N_p>25$\hfill \\
\enddata
\label{tab:sigmar}
\end{deluxetable}

\section{Discussion and Conclusions}
\label{sec:disc}
In the previous section we presented results of numerical
simulations with initial conditions constructed using large-scale
constraints from the observed density field (deduced using the \m3\ 
peculiar velocity survey) and a random realization of the density field
for the popular \LCDM\ model of structure formation.  These initial
conditions lead to the formation of large-scale ($\gtrsim 5$\Mpch)
structures that match the observed structures within $\sim
100\hmpc$ around the Local Group well, while the structures on smaller
scales are not constrained and represent a random realization of the
\LCDM\ primordial power spectrum. 

We show that such {\em constrained simulations}  reproduce the
most prominent nearby large-scale structures, such as the Coma
cluster, the Great Attractor, the Perseus-Pisces supercluster, the
LSC, and the Local Void.  The locations of the
structures varies within the ensemble of the constrained simulations (with different
random realizations of the \LCDM\ spectrum) by a few megaparsecs, but
the overall qualitative large-scale morphology is reproduced robustly.
The differences in structure location from realization to realization
imply that the constrained simulations should not be expected to reproduce the observed
structures accurately in the highly non-linear regime.  Instead, they
should be considered as a tool for generating realizations of the
primordial density field for specific purposes and comparisons with
observations in which it is crucial that the cosmic variance is
greatly reduced. In other words, constrained simulations are well suited for studying
the formation of particular objects, density field configurations, and
environments. The outputs of constrained simulations can be used for generation of mock
galaxy catalogs and testing empirical models (e.g., the peculiar
velocity field models).

In this paper we used the constrained simulations (described in detail in \S~\ref{sec:CS}) 
to study the morphology of the density distribution and the peculiar 
velocity field in the $\sim 30\hmpc$ region around the Virgo cluster 
(which we call the LSC region in this study).  In particular, we 
focused on the question of the relative ``coldness'' of the peculiar 
velocity field in the immediate vicinity of the LG. We find that the 
velocity field in the LSC region is quite complex.  Matter and halos 
in underdense regions, such as the Local Void, undergo coherent 
expansion as predicted by analytic models of void evolution 
\citep{hoffman_shaham_82,bert85}.  Near large-scale filaments and 
sheets the peculiar velocities are nearly perpendicular to the longest 
dimension of the closest structure, the pattern predicted by the 
\citet{zeldovich70} collapse solution.

The main body of the LSC (the $\sim 20\hmpc$ filament centered on the
Virgo cluster) in the simulation contains most of the mass in the LSC
region ($\approx 7.5\times 10^{14}\hmpc$), while the LG is
located in an inconspicuous filament bordering the Local Void. The
virial mass of the Virgo cluster is $\approx 1\times 10^{14}\hmpc$ or
less than 15\% of the LSC mass and therefore the cluster does not
dominate dynamics around the LSC. Correspondingly, a spherically
symmetric Virgocentric infall model provides a very poor description of
the peculiar velocity field which shows coherent infall in the overall
direction of the extended mass concentration of the LSC rather than
towards the Virgo cluster. In particular, the model overestimates the
mass within $10\hmpc$ around the Virgo cluster in our simulation by a
factor of three. The value of matter density $\Omega_0$ inferred using
this model is thus grossly inaccurate. This inaccuracy was pointed out
in previous tests of this model in the LG like environments
in cosmological simulations \citep{cen94,governato_etal97}.

The matter within and around the LG counterpart in our
simulation participates in the infall onto the LSC (roughly in the
direction of the Virgo cluster) with a velocity of $\sim 250{\ \rm km\ 
  s^{-1}}$.  Although the observational measurements of the LG infall
velocity span a wide range of values, the average value, $\sim 250\pm
50~\kms$ \citep[see, e.g.,][ and references therein]{huchra88}, is in
good agreement with that found in our simulation. The local overdensity of
dark matter around the simulated LG is $\approx 5.5$ within
$1.5\hmpc$ and is about zero within $5\hmpc$. This can be compared to
the overdensity of galaxies of $\sim 0.25$ within $5\hmpc$ in the IRAS
survey \citep{schlegel_etal94}. Given the small statistics and
observational uncertainties this is a fair agreement: both the
simulated and real LG appear to reside in only slightly
overdense environments.

Although the LG is located in a filament, this filament is of
relatively small mass and does not perturb the large-scale infall flow
significantly.  Therefore, the flow in the immediate vicinity of the
LG is rather smooth. This explains the relative ``coldness'' of the
local peculiar velocity field.  The radial velocity dispersion within
$\approx 7\hmpc$ around the LG in the simulation is $\sim
40-70~\kms$, which is in good agreement with
observations \citep[e.g.,][ and references
therein]{sandage86,giraud86,schlegel_etal94,km96,ekholm_etal01}.  The
velocity dispersion increases to $\approx 110{\ \rm km\ s^{-1}}$ at
larger distances due to larger peculiar velocities within and around
the main body of the LSC.

The low value of the velocity dispersion we find in our simulation is
in contrast with considerably higher values found for the simulated LG
like systems in previous studies
\citep{schlegel_etal94,governato_etal97}. We are not certain what
caused the differences.  We think that some of the differences are due
to the different cosmological model. Very likely this is so in the
case of the SCDM model as opposed to \LCDM~ model studied in this
paper.  In case of the open CDM model studied by
\citet{governato_etal97}, which has the same density of matter as in
our \LCDM~ model, the differences may also be partially attributed to
a more accurate representation of the large-scale environment of the
LG in our simulation.  In any case, it is comforting that the ``cold''
local velocity field is reproduced within the current favored {\LCDM}
cosmological model and there is no need for exotic explanations
\citep[e.g.,][]{chernin_etal01}.

To conclude, we think that numerical experiments conducted under
controlled conditions of constrained realizations should help to shed
light on some of the unanswered questions in the theory of structure
and galaxy formation. Some of these questions arise from the unique
data in the immediate neighborhood of the LG.  The observational
constraints on the primordial density field can be strengthened with
the improving size and accuracy of the peculiar velocity surveys.
Indeed, even in the immediate future one can conceivably perform more
accurate constrained simulations by combining the \m3\ survey used in
this study and the Surface Brightness Fluctuations
\citep[SBF;][]{tonry_etal01} peculiar velocity survey, which has more
accurate velocities for nearby galaxies.

The question of the coldness of the local peculiar velocity field
addressed in this paper is but a single example of the problems that
can be addressed by constrained simulations.  Indeed, in a related
paper we extend the present work by using gas-dynamics constrained
simulations to study properties of the intergalactic medium in the LSC
region and formation of the Virgo cluster \citep{kkh01}. We also plan
to use higher mass and force resolution constrained simulations to
study the distribution of dwarf galaxies in the LSC and nearby voids,
formation of the LG and other nearby groups and clusters,
etc. Overall, we find that the constrained simulations provide an
optimal tool for studying a wide range of dynamical problems
concerning our ``local'' neighborhood within the large-scale structure
of the Universe.

\acknowledgements We thank Rien van de Weygaert for discussions during
initial stage of the project.  Y.H. has been supported in part by the
Israel Science Foundation (103/98).  We acknowledge support from the
grants NAG- 5- 3842 and NST- 9802787 to NMSU.  A.V.K. was supported by
NASA through Hubble Fellowship grant from the Space Telescope Science
Institute, which is operated by the Association of Universities for
Research in Astronomy, Inc., under NASA contract NAS5-26555. SG
acknowledges support from Deutsche Akademie der Naturforscher
Leopoldina with means of the Bundesministerium f\"ur Bildung und
Forschung.  AAK thanks the Institute of Astronomy at Cambridge for
hospitality and support during his visit.  Computer simulations
presented in this paper were done at the National Center for
Supercomputing Applications (NCSA) at Urbana-Champaign and at the
Leibnizrechnenzentrum (LRZ) in Munich.
\bibliographystyle{apj}

\bibliography{lsc}

\end{document}